# Proactive Empirical Assessment of New Language Feature Adoption via Automated Refactoring: The Case of Java 8 Default Methods

Raffi Khatchadourian[a,b] and Hidehiko Masuhara[c]

a   Hunter College, City University of New York, USA
b   Graduate Center, City University of New York, USA
c   Tokyo Institute of Technology, Japan

**Abstract**   Programming languages and platforms improve over time, sometimes resulting in new language features that offer many benefits. However, despite these benefits, developers may not always be willing to adopt them in their projects for various reasons. In this paper, we describe an empirical study where we assess the adoption of a particular new language feature. Studying how developers use (or do not use) new language features is important in programming language research and engineering because it gives designers insight into the usability of the language to create meaning programs in that language. This knowledge, in turn, can drive future innovations in the area. Here, we explore Java 8 default methods, which allow interfaces to contain (instance) method implementations.

Default methods can ease interface evolution, make certain ubiquitous design patterns redundant, and improve both modularity and maintainability. A focus of this work is to discover, through a scientific approach and a novel technique, situations where developers found these constructs useful and where they did not, and the reasons for each. Although several studies center around assessing new language features, to the best of our knowledge, this kind of construct has not been previously considered.

Despite their benefits, we found that developers did not adopt default methods in all situations. Our study consisted of submitting pull requests introducing the language feature to 19 real-world, open source Java projects without altering original program semantics. This novel assessment technique is proactive in that the adoption was driven by an automatic refactoring approach rather than waiting for developers to discover and integrate the feature themselves. In this way, we set forth best practices and patterns of using the language feature effectively earlier rather than later and are able to possibly guide (near) future language evolution. We foresee this technique to be useful in assessing other new language features, design patterns, and other programming idioms.

# The Art, Science, and Engineering of Programming

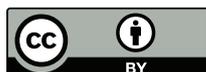

## 1 Introduction

Programming languages and platforms change for a variety of reasons such as to accommodate new environments and/or trends or to add new language features. For example, new Java features have included generics and lambda expressions [8, 9].

To benefit from new language features, however, developers must be willing to adopt them into their projects. Developers may or may not favor a certain feature, and several researchers have investigated the reasons. For instance, Dyer, Rajan, Nguyen, and Nguyen [14] and Parnin, Bird, and Murphy-Hill [31] investigate how (or if) generics and other Java features have been adopted by mining open source software. Uesbeck, Stefik, Hanenberg, Pedersen, and Daleiden [40] assess the impact of C++ lambda expressions on developers' experience by analyzing log data, and Wu, Chen, Zhou, and Xu [41] analyze C++ concurrency construct usage in open source software. Gorschek, Tempero, and Angelis [18] and Tempero, Counsell, and Noble [38] study developers' usage patterns of Object-Oriented concepts like encapsulation and inheritance via surveys and metrics, respectively.

This paper describes an empirical study where we assess the adoption of a particular new language feature, namely, default methods, part of Java 8's *enhanced* interfaces. Default methods enable developers to write *default* (instance) methods that include an implementation that implementers will inherit if one is not provided [8]. Its original motivation was to effectively facilitate interface evolution, i.e., interface augmentation without breaking existing clients [7]. Default methods can also be used [17, 24] as a substitution for the *skeletal implementation* pattern, which employs an (abstract) skeletal implementation class that interface implementers extend [3, Item 18]. In this way, interfaces are easier to implement due to the class providing a partial interface implementation.

Although default methods can be useful in making interfaces more evolvable and bypassing the need for auxiliary design patterns, during our study, we found that there are also trade-offs to using default methods that may not be obvious. For example, certain implementations may be considered too general to serve as a default implementation to be inherited by all interface implementers, which is in contrast to the skeletal implementation pattern as these would be available to them regardless of any subclassing. Also, interface authors lose the guarantee that *all* clients will provide their own specific implementation of a method. Moreover, adding non-trivial implementations to interfaces may result in module dependencies that violate existing architectural constraints.

In the sections that follow, we detail reactions of developers in adopting default methods in their projects, extracting best practices of their uses, as well as situations where these new constructs work well and where trade-offs must be made by performing an empirical study on 19 real-world, open source Java projects hosted on GitHub.[1] Pull requests (patches) were issued that contained particular interface method im-plementations migrated to interfaces as default methods in a semantics-preserving fashion.

---

[1] http://github.com, last accessed on March 23, 2018.

A popular approach for assessing language features, as utilized by related work (e.g., [14, 28, 29, 31, 40, 41]), involves a *postmortem* analysis. Specifically, either *past* data (e.g., source code, commits) of source repositories are analyzed or surveys of previous coding activities are taken. This approach has several problems. Firstly, developers must discover new language features and integrate them themselves before any analysis of the construct can be done. The necessary time for this to take place can be extremely long and new language versions may not benefit from the analysis prior to the next release. Best practices and patterns that can normally be extracted from these studies are also delayed. Furthermore, developers may be unable to *manually* identify *all* opportunities where the new language construct can be utilized. Lastly, observing software histories may discover cases where new language features are *adopted* but may not easily identify those where they were *rejected* as these may not have been adequately documented.

To combat these problems, we introduce a novel technique for assessing new language constructs *proactively*. The pull request changes in our study consist of transformations performed via an automated refactoring tool. As such, developers are immediately introduced to the new construct, irrespective of any previous experience, via a semantically equivalent transformation that they can either accept or reject. Their decisions can be studied early to assess the feature's effectiveness, extracting best practices.

The use of refactoring automation is key in minimizing human bias that such a proactive approach can introduce. Furthermore, it is also extremely important that the tool be highly conservative in its refactoring approach, as well as founded in a firm theoretical basis. In this way, the refactoring tool performs transformations that are (i) correct (i.e., semantics is preserved), (ii) minimally invasive (i.e., composite refactorings are not performed so that singular construct under question can be studied in isolation), and (iii) a subset of those that a project owner would have performed.

To this extent, we chose to use the Migrate Skeletal Implementation to Interface refactoring tool [23], based on type constraints [30, 39], to conservatively and confidently discover opportunities and necessary semantics-preserving transformations for migrating methods possibly participating in the skeletal implementation pattern to interfaces as default methods for this study. Its conservative nature and use of a well-founded type constraint framework ensures that project owners can be confident that it makes correct changes, that it only makes the changes necessary to introduce default methods into existing projects, and only makes changes in unambiguous situations, i.e., where there is only one possible outcome. In other words, the approach is fully-automated.

Our study assesses the use of default methods in *existing* code. In other words, as previously mentioned, the original motivation for default methods was interface extension, i.e., adding *new* methods to an interface, without breaking existing clients [17]. Substituting the skeletal implementation pattern is the only sensible use of default methods when not introducing new functionality. Thus, an acceptance of the refac-toring results in this context for reasons pertaining only to the result content itself is equivalent to acceptance of using default methods as a programming construct for existing code and vice-versa.

Our approach is highly appropriate for studying the adoption of default methods as Java 8 is still relatively new. Also, the refactoring tool can *automatically* locate feature usage opportunities that developers may have missed otherwise, allowing us to study more occurrence where the construct may be used. Finally, pull request rejections were typically accompanied by the thorough and thoughtful reasoning of why the language feature was not to be used. This would have otherwise not been documented or been easily identifiable in the source code as developers may subsume such reasons implicitly.

Preliminary results of this pull request study are briefly presented by Khatchadourian and Masuhara [23], however, its sole focus is evaluating the usability of the refactoring approach itself as a secondary metric. In other words, the discussion there argues that the tool results are *useful* in real-world scenarios and is *not* intended to study the construct's applicability. That, though, is precisely one of the main goals of this work. Here, our intent is not to evaluate the refactoring tool itself but rather describe key characteristics of why developers either accepted or rejected using default methods. Doing so is possible due to the above discussion regarding the tool's conservative qualities.

From these observations, we additionally set forth best practices for default method use, identifying where they are most effective, as well as where and when they should be avoided. The results provide valuable insight into the applicability of this new construct to real software, pinpoint their best usage scenarios, including those where default methods may not be, perhaps surprisingly, advantageous, and guide future language evolution, for Java or other strongly-typed languages considering similar constructs.

This paper makes the following contributions:

**Proactive Evaluation**  Our study features a novel empirical analysis technique to assess new language features early rather than later by using an automated refactoring tool to introduce developers to the new construct regardless of their previous experience. Developers are informed that the transformation is semantic-preserving. Moreover, this facilitates the study of otherwise possibly unobservable reasons for unadopted cases.

**Practicality Assessment**  Key insights into the adoptability of default methods in 19 real-world, vetted open source projects of varying size, domain, and popularity are given. Scenarios where default methods made sense and where it did not and why are thoroughly detailed. This highlights the practicality of default methods in existing code.

**Best practices Extraction**  From these scenarios, we describe general best practices for using default methods. These practices (or patterns) are extracted from real software. Carefully considered evidence is presented to support the claims.

**Listing 1** An example migration from the skeletal implementation pattern to default methods. Inspired by [10, 23, 39].

**(a)** Abstract skeletal implementation pattern.   **(b)** Migration to default methods.

```
 1  interface List<E> {
 2    int size();
 3    boolean isEmpty();
 4    int capacity();
 5    abstract boolean atCapacity();
 6
 7    void setSize(int i);
 8    void removeLast(); // optional operation.
 9
10    void set(int i, E e);
11    void print(PrintStream stream);}
12  abstract class AbsList<E> implements List<E> {
13    @Override public boolean isEmpty()
14      {return this.size() == 0;}
15    @Override public boolean atCapacity()
16      {return this.size() == this.capacity();}
17    @Override public void removeLast()
18      {throw new UnsupportedOperationException();}
19    @Override public void print(PrintStream out)
20      {out.println(this);}}
21  class ArrayBasedList<E> extends AbsList<E> {
22    Object[] elems; int size; // instance fields.
23    @Override public int size() {return this.size;}
24    @Override public void setSize(int i) {this.size = i;}
25    @Override public int capacity()
26      {return this.elems.length;}
27    @Override public void set(int i, E e)
28      {this.elems[i] = e;}}
29  class Main {
30    public static void main(String[] args) {
31      List<Integer> list = new ArrayBasedList<Integer>();
32      assert(list.isEmpty());
33      assert(!new AbsList<String>() {/*...*/}.atCapacity());}}
```

```
 1  interface List<E> {
 2    int size();
 3    default boolean isEmpty() {return this.size() == 0;}
 4    int capacity();
 5    default boolean atCapacity()
 6      {return this.size() == this.capacity();}
 7    void setSize(int i);
 8    default void removeLast() // optional operation.
 9      {throw new UnsupportedOperationException();}
10    void set(int i, E e);
11    default void print(PrintStream out) {out.println(this);}}
12  abstract class AbsList<E> implements List<E> {
13    @Override public boolean isEmpty()
14      {return this.size() == 0;}
15    @Override public boolean atCapacity()
16      {return this.size() == this.capacity();}
17    @Override public void removeLast()
18      {throw new UnsupportedOperationException();}
19    @Override public void print(PrintStream out)
20      {out.println(this);}}
21  class ArrayBasedList<E> implements AbsList<E> {
22    Object[] elems; int size; // instance fields.
23    @Override public int size() {return this.size;}
24    @Override public void setSize(int i) {this.size = i;}
25    @Override public int capacity()
26      {return this.elems.length;}
27    @Override public void set(int i, E e)
28      {this.elems[i] = e;}}
29  class Main {
30    public static void main(String[] args) {
31      List<Integer> list = new ArrayBasedList<Integer>();
32      assert(list.isEmpty());
33      assert(!new AbsList<String>() {/*...*/}.atCapacity());}}
```

**Organization**

The remainder of this paper is organized as follows. Default method and skeletal implementation pattern concepts are explained in Section 2. Section 3 discusses our experimental methodology, its setup, etc. In Section 4.1, details of the issued pull requests are described. Section 4 presents the results of the study, specifically, a depiction of the developers reactions to the pull requests. A discussion of the results resides in Section 5, along with extracted patterns and best practices. Section 6 presents the threats to validity. Section 7 compares related work, and Section 8 concludes and details plans for future work.

## 2 Background

## 2.1 Skeletal Implementation Pattern

While default methods are useful for introducing new methods into interfaces without breaking existing clients, in this study, default methods are integrated into projects via a semantics-preserving transformation (refactoring) of methods possibly participating in the widely-used [23] skeletal implementation pattern. To illustrate what a developer may be presented with as part of the study, Listing 1 portrays a hypothetical collection type hierarchy snippet inspired by related work [10, 23, 39]. For presentation purposes, the hierarchy has been simplified, showing only relevant portions. The original system (white space added for alignment) is pictured in Listing 2a, while Listing 2b depicts the same system with several methods migrated to interfaces as default methods. Removed code is struck through, added code is underlined, and replaced code is both underlined and emphasized. Both systems are type-correct and semantically equivalent. We refer readers to Khatchadourian and Masuhara [23] for more details of the refactoring.

A List interface is shown on lines 1–11. Note that, unlike classes, Java interfaces can extend multiple interfaces. Methods exist for determining a List's size(), whether it isEmpty(), its capacity(), whether it is atCapacity(), setting a List's size, removing its last element, replacing an element at a specified position, and printing it to a specified stream. removeLast() is denoted as a so-called *optional* operation as, e.g., not all list types may support deletion. In such cases, implementers may throw an exception when these methods are invoked.

AbsList, an abstract class providing a skeletal implementation of a List, is declared on line 12. It "assists" classes with the interface implementation by declaring appropriate basic method implementations for the more primitive operations. Since it is abstract, it is not required to implement all interface methods. For the optional operation removeList(), the provided implementation (line 18) simply throws an Unsupported-OperationException. This way, concrete implementers extending AbsList that support element removal can override it with a working implementation, while others need not override it. The provided implementation of print() effectively sends the standard string representation of the List (the result of Object.toString()) to the stream. Notice that all method implementations besides print() invoke only methods contained within the interface; print() includes a call to the println() method that resides outside of the hierarchy.

ArrayBasedList (lines 21–28) is a sequential, variable length List implementation. Main is a driver whose main method (line 30–33) instantiates a concrete List, as well as a subclass of the AbsList skeletal implementations classes via an anonymous inner class (AIC, line 33).

Listing 2a illustrates the skeletal implementation pattern. As this example depicts, there are several drawbacks, which include inheritance restrictions, e.g., ArrayBasedList must extend the skeletal implementation class and cannot easily take advantage of other skeletal implementations if it were to extend multiple interfaces [20].[2] Furthermore, a whole program analysis may be required to discover skeletal implementation classes as there is no syntactic path between them and the interface, adversely affecting modularity [24]. Although modern IDEs may help alleviate this problem, reliance on IDEs can be time consuming and costly as they can be resource-heavy [4]. Also, the pattern may result in bloated libraries due to the additional classes needed, possibly making maintenance challenging, especially since method declarations (in the case of classes, method headers) must be repeated in the skeletal implementation class.

**2.2 Default Methods in Java 8**

Default methods enable skeletal implementations in interfaces, thereby foregoing the need to place such methods in separate classes. Moreover, interface implementers need not search for separate skeletal implementations classes. Lastly, implementers are free to extend classes other than the skeletal implementation class, as well as inherit behaviors from *multiple* interfaces, which can reduce code duplication and forwarding methods [17].

Listing 2b shows a refactored version of the code in 2a. All skeletal implementations in `AbsList` have been migrated to `List` as default methods. To later facilitate discussion in Section 4, we denote methods that are being migrated to interfaces as *source* methods, the class declaring the source method as the *declaring class*, the interface in which the source methods are migrated to as their corresponding *destination* interface, and the (non-default) interface methods they replace (i.e., those that become default methods) as *target* methods (cf. Table 2). Note it is possible for a class to contain source methods with different destination interfaces and for target methods to have multiple source methods [23].

`ArrayBasedList` (Listing 2b) now implements `List` rather than extending `AbsList` (line 21); it can now extend other classes. The call at line 32 now dispatches to the interface implementation, which is equivalent to the one formerly belonging to the (now removed) abstract class. Since `AbsList` no longer exists as a result of the refactoring, line 33 is updated to be an AIC directly implementing the interface.

Now, developers considering implementing `List` can clearly recognize the default behavior for methods without consulting a separate class or documentation. For instance, it is clear that `List.removeLast()` is optional and what should happen when it is not implemented. Classes can simply inherit default implementations without needing to find and subclass a separate skeletal implementation class or, perhaps worse, duplicate existing code.

---

[2] Implementers already extending a class can use the skeletal implementation via delegation to an internal class [3] at the expense of auxiliary forwarding code.

- **Table 1** Pull requests.

| | subject | pull ID | KLOC* | watches† | stars† | forks† | contribs† | +LOC | -LOC | δ files | concrete? |
|---|---|---|---|---|---|---|---|---|---|---|---|
| merged | aalmiray/jsilhouette | 1 | 2 | 2 | 4 | 1 | 2 | 147 | 294 | 4 | *false* |
| | aol/cyclops-react | 258 | 99 | 68 | 554 | 54 | 21 | 8 | 15 | 2 | *false* |
| | eclipse/eclipse-collections | 128 | 1,266 | 40 | 258 | 63 | 18 | 172 | 307 | 21 | *false* |
| | nhl/bootique | 79 | 5 | 103 | 744 | 183 | 5 | 22 | 31 | 4 | *true* |
| rejected | iluwatar/java-design-patterns | 472 | 20 | 1,783 | 17,234 | 5,808 | 71 | 24 | 38 | 6 | *false* |
| | jOOQ/jOOQ | 5469 | 136 | 127 | 1,614 | 411 | 40 | 93 | 187 | 22 | *false* |
| | google/guava | 2519 | 244 | 1,568 | 14,721 | 3,502 | 98 | 241 | 427 | 16 | *false* |
| | google/binnavi | 99 | 309 | 215 | 2,048 | 373 | 16 | 244 | 469 | 16 | *false* |
| | eclipse/jetty.project | 773 | 329 | 196 | 1,225 | 811 | 61 | 140 | 263 | 29 | *false* |
| | spring-projects/spring-framework | 1113 | 506 | 2,299 | 12,463 | 9,575 | 200 | 770 | 1,674 | 135 | *false* |
| | elastic/elasticsearch | 19168 | 1,266 | 1,928 | 21,063 | 7,275 | 784 | 297 | 544 | 51 | *false* |
| | jenkinsci/blueocean-plugin | 296 | 7 | 114 | 1,688 | 173 | 28 | 8 | 19 | 5 | *true* |
| | junit-team/junit5 | 5365 | 25 | 146 | 865 | 215 | 41 | 4 | 18 | 1 | *true* |
| | ReactiveX/RxJava | 4143 | 154 | 1,677 | 21,792 | 3,819 | 142 | 29 | 131 | 23 | *true* |
| pending | perfectsense/dari | 218 | 66 | 111 | 48 | 31 | 28 | 39 | 58 | 7 | *false* |
| | eclipse/jgit | 34 | 172 | 57 | 429 | 247 | 121 | 35 | 127 | 10 | *false* |
| | rinfield/java8-commons | 81 | 2 | 1 | 0 | 2 | 1 | 26 | 48 | 3 | *true* |
| | criscris/koral | 1 | 7 | 1 | 1 | 1 | 1 | 169 | 197 | 6 | *true* |
| | advantageous/qbit | 767 | 52 | 82 | 534 | 115 | 12 | 80 | 202 | 29 | *true* |
| | Totals: | | 4,665 | 10,518 | 97,285 | 32,659 | 1,690 | 2,548 | 5,049 | 390 | |

* At time of analysis.
† As of February 27, 2017.

## 3 Methodology

The refactoring described in Section 2 is used to present developers with modifications (pull requests) of their projects that utilize default methods. It conservatively and confidently discovers opportunities and necessary semantics-preserving transformations for migrating methods possibly participating in the skeletal implementation pattern to interfaces as default methods. It does so unambiguously, i.e., the refactoring tool only takes action when there is only one possible choice to be made, requiring no input from the developer. Because of this, the presented study is that of the construct rather than the tool's performance. We then study the adoption of default methods into developers' projects.

### 3.1 Research Questions

We seek answers to the following research questions:

R1. In which situations do developers adopt default methods in their projects? What are the reasons?
R2. Despite their benefits, are there situations where developers do not favor default methods?
R3. What are the trade-offs of using default methods over the skeletal implementation pattern?
R4. Which external factors, if any, influence developer's decisions in adopting default methods?
R5. Are there best practices and/or patterns that can be extracted from these situations?

### 3.2 Refactoring Tool Implementation

The automated refactoring tool MIGRATE SKELETAL IMPLEMENTATION TO INTERFACE [23][3] is implemented as an open source plug-in for the Eclipse IDE[4] and is built upon an existing refactoring framework [2]. Eclipse ASTs with source symbol bindings are used as an intermediate representation.

### 3.3 Subjects

19 open-source Java applications and libraries were selected for the study (Table 1). We purposely selected projects that had the majority of their code base in Java and were not Android projects since Java 8 was to supported in Android at the time of the study. The projects were also selected so that they range in size, domain, and popularity. Both projects that had moved to Java 8 and those that had not were used in the study. For projects that had not yet moved to Java 8, our pull request typically

---

[3] Available at http://git.io/v2nXo, last accessed March 23, 2018.
[4] http://eclipse.org, last accessed March 23, 2018.

contained modifications to the build script to increment the Java version number to use.

Subjects comprised projects (column *subject*) from organizations such as Eclipse, AOL, and NHL (National Hockey League), and include the Eclipse (formally Goldman Sachs) Collections framework. Column *pull ID* is the pull request ID for the associated project.[5]

Column *KLOC* is the number of non-blank, non-comment thousands of lines of Java source code[6] at the time the analysis was performed. Columns *watches*, *stars*, *forks*, and *contribs* denote the number of GitHub users monitoring the project, number of users that marked the project as a "favorite," the number of users that copied the project into their own personal space for modification, and the number of individual project contributors, respectively. Such attributes give insight into the popularity and usage of the subjects. More details regarding the subject projects may be obtained from the respective GitHub pages.[7] The remaining columns are discussed throughout the sections that follow.

### 3.4 Pull Requests Issuance

**Refactoring Application**   To study developer reactions to the introduction of default methods, pull requests were submitted to each of the projects listed in Table 1. To minimize experimental variability, we ensured that projects compiled correctly and had identical test results and compiler warnings before and after the refactoring.[8] The MIGRATE SKELETAL IMPLEMENTATION TO INTERFACE refactoring tool automatically analyzes and transforms, where applicable (i.e., where refactoring preconditions passed), each project to utilize default methods in lieu of the skeletal implementation pattern. It mines for occurrences of the pattern and then determines instances of which that (i) are viable candidates for the refactoring (as dictated by language constraints), (ii) would result in a semantics-preserving transformation, and (iii) have one and only one possible destination.

**Manual Intervention**   In some cases, minor manual cosmetic intervention was required for the delta to more closely resemble changes made by a human. Such changes included the merging of documentation (Javadoc) between method definitions in classes and those in interfaces. These changes did not have a direct effect on the acceptance of default methods but do make the submitted pull requests more amenable to acceptance. However, there were other manual changes that were necessary, specifically, to fix build (e.g., Maven [16], Gradle [22]) module dependencies for source methods that crossed module boundaries, which was out of the scope of the automated refactoring. This is discussed further in Section 4.2.4. Whether refactoring changes are completed or mostly automated w.r.t. artifacts other than source code should not influence developers' final decisions.

---

[5] Pull request URLs consist of http://github.com/, the *subject* column value, /pull/, and the pull request ID.
[6] Generated using 'SLOCCount' by David A. Wheeler.
[7] Project URLs are http://github.com/ followed by the *subject* column value.
[8] Running tests locally prior to issuing a pull request was sometimes complicated by differences in test machine environments, resulting in pull request reissuance.

**Convincing Developers of Semantics-Preservation** Most subjects used continuous integration (CI), and an effort was made to apply the refactoring to the head of a previous commit that had passing tests. To gain developers' confidence that the proposed refactorings were indeed semantics-preserving, since CI would run the projects' test suite on our commits, we ensured that the tests either passed or had the same results as the previous commit. Importantly, our commits did not contain any significant changes to test code.

For projects that did not include CI, the test suite was run locally, and the test results were conveyed to the developers via pull request messages. For projects that did not include a (working) test suite, developers analyzed change sets to ensure semantics-preservation.

## 4 Results

### 4.1 Pull Requests and Change Sets

In Table 1, columns *+LOC* and *-LOC* depict the number of added and removed lines of code, respectively, in the pull request, while column *δ files* is the number of changed files. These statistics are from the git commits comprising the pull requests. Changes ranged from large, with the largest submitted to spring-projects/spring-framework, totaling 2,444 changed lines of code and 135 files, to small, with one of the smallest stemming from junit-team/junit5, totaling 22 changed lines and 1 file.

These columns give insight into the scope of the changed proposed to each subject project; naturally, one would expect that large, more pervasive change proposals would be treated more skeptically. In fact, among pull requests with changes to $\geq 10$ files, only one project, namely, eclipse/eclipse-collections, merged (accepted) the request. On the other hand, projects with requests consisting of $< 10$ changed files were equally likely to be merged. This result is statistically significant in that the associated p-value on the number of changed files against merged and rejected requests from a Mann–Whitney $U$ test, calculated using R [32], is 0.05156[9] (a value $< 0.1$ is considered statistically significant). The script used to calculate p-values is available on our website.[10]

During the study, it was found that developers did not always use the classic version of the skeletal implementation pattern. Particularly, would-be skeletal implementation classes were not always abstract, though, they seemed to follow much of the essence of the pattern. This was particularly the case with smaller projects and those not necessarily providing APIs for use by other applications. Other times, we did not wish to prevent a project from participating due to its small number of classes. In other words, although a project may not follow the pattern exactly, there may still be methods that pass refactoring preconditions for migration. To accommodate these projects, the refactoring tool was set to allow source methods declared in concrete classes. Typically, abstract classes serve as declaring classes in the classical version of the pattern.

---

[9] A one-tailed test is used since the hypothesis is directional.
[10] http://cuny.is/interefact, last accessed March 23, 2018.

Column *concrete* is true iff there exists at least one source method declared in a concrete class. It was slightly more likely for a pull request containing concrete methods to be rejected; only 25 % of such non-pending requests were merged as opposed to 30 % that only contained source methods declared in abstract classes, having a p-value of 0.4641. Indeed, only one project, nhl/bootique, merged a request that contained such methods.

Table 1 is divided into three distinct row-wise sections, *merged*, *rejected*, and *pending*. A merged pull request is one where the refactoring results have been integrated into the project, a rejected request is one where developers refused to incorporate the refactoring results, and pending requests are ones where developers' decisions have not been finalized.

Of the issued 19 pull requests, 4 were successfully merged, 10 were closed without merging (rejected), and 5 are still pending (open). Projects with merged requests averaged ∼343 KLOC, slightly higher than the average of those projects that did not (∼299 KLOC), with a p-value of 0.1608. However, although several of the projects with merged requests were popular, other project attributes, e.g., stars, for projects with rejected requests were, on average, much higher than those of the merged requests, having an average p-value of 0.001998. This may indicate that more widely used projects were more risk-adverse than less used ones.

Likewise, the average pull request change set size, both in terms of LOC and number of changed files, was significantly greater for rejected requests than for those that were merged, indicating that pull requests consisting of larger change sets were more likely to be rejected. However, the p-value for changed files is 0.05156, but the p-value for total changed LOC is 0.2697. Thus, the former is statistically significant, while the latter is not. Nevertheless, two merged pull requests, namely, those for the aalmiray/jsilhouette and eclipse/eclipse-collections, had comparably sized change sets to some of the larger sets of rejected requests. Moreover, the total number of "changed" lines (added LOC plus the removed LOC) for the merged request sent to aalmiray/jsilhouette was 441, which is large compared to its total code base, specifically, 19.33 %. This merged pull request consisted of the largest percentage of change proposed to any one project, with the next largest at 5.14 % for criscris/koral.

Rejected pull requests were not unequivocally due to the the contents of the refactoring results. Some projects, e.g., google/guava, had external factors like being in the middle of release cycles and desired to consider the pull request at a later time. In fact, at the time of our request issuance, google/guava was in the process of creating a Java 8 branch of their project, and the introduction of our automated refactoring may have complicated their migration that was already in progress.

◼ **Table 2** Refactoring terminology.

| Term | Description |
| --- | --- |
| Source Method | The class instance method being migrated to an interface as a default method. |
| Declaring Class | The class declaring the source method. |
| Target Method | The interface (abstract) method that will be converted to a default method. |
| Destination Interface | The interface declaring the target method. |

## 4.2 Observations

We now detail several categories and particular instances of when pull requests were either accepted or rejected.

### 4.2.1 Backwards Compatibility with Earlier Java Versions

Several projects, including ReactiveX/RxJava and google/binnavi, rejected our pull requests, citing reasons such as that they had not yet moved (or were in the process of moving) to Java 8 and/or needed to support older Java clients at the time. Particularly, many such projects were required to support Android clients. At the time of this study, Android had not officially fully supported all Java 8 features (but does support at least default methods currently [21]).

### 4.2.2 Client Impact and Loss of Implementation Control

**Requiring Explicit Client Implementations** As discussed in Section 2.2, interface implementers will inherit the default implementation of an interface method if they do not provide (or inherit) their own. In fact, this is one of the main benefits of the Skeletal Implementation Pattern, i.e., interface implements will receive interface method implementations that are common to all implementers. In the case of optional methods, interface implementers need only provide implementations for the methods they support; the default implementation could just throw a UnsupportedOperationException, as shown in Listing 1, or simply be blank. We found several instances of the latter in pull requests issued to projects such as ReactiveX/RxJava. Part of the reason these pull requests were rejected was because developers preferred to force clients to implement particular interface methods. Converting these methods to default methods would not have this effect, i.e., interface developers would lose control of interface clients providing their own implementations for particular methods. This was particularly noticed in the case of what we thought were optional methods as the skeletal implementations consisted of an empty body. What was surprising is that the refactoring introduced only *minimal* code, and, as such, the (undesirable) skeletal implementations emanate from an *existing* corresponding skeletal implementation class. Moreover, the refactoring tool is very conservative and only performs the migration if one and only one valid source/target method combination exists (i.e., there are no ambiguous target methods for a single source method) [23]. As such, a more suitable skeletal implementation should not have existed in the system at the time.

**Influencing Client Implementations**   In the case of elastic/elasticsearch, the developers agreed that no suitable "default" implementation existed for their highly utilized `Client` interface, that all clients of their framework implement. This is *despite* the project providing an abstract skeletal implementation class `AbstractClient`. From the developer's comments, we speculate that separating default implementations from interfaces was desirable so that only seemingly savvy implementers would find them. Such implementers may be more aware of the ramifications of utilizing a common behavior for all clients.

> Tedor [37] states in response to the migration of skeletal implementations to default:
>
>> I'm not sure if we should do this, this assumes behavior on all `Client` implementations (which exist outside of core elasticsearch). `AbstractClient`, though, is exactly for those that do want the behavior here.
>>
>> I think of **default** as "it's okay if implementers do not implement this method because this default implementation should work for all of them if they maintain the invariants assumed by this interface" but I don't think that we can safely say that that is the case here.

We observe that placing default implementations in *core* packages, where many interface declarations reside, would advocate or even encourage certain implementations. On the contrary, having such implementations *outside* the interface would deter from this notion.

**Interface Evolution**   There was also a concern in elastic/elasticsearch that the "default methods may cause ... implementations to miss new methods" [5]. In effect, one reason this project was leery was for the fear that upon interface evolution, existing clients would *not* break. This breakage is desirable because when linking against a new framework/library version, clients would be required to implement new methods to compile correctly. Default methods may "hide" the fact the new interface methods exist that need specific (new) implementations from existing implementers. Projects, in this way, use the compiler to "announce" to implementers that interface functionality has been extended. Providing a default implementation may result in these new methods going unnoticed.

**Risk vs. Reward**   Although intrigued by the refactoring, project eclipse/jetty.project ultimately, after much discussion, rejected our pull request. The developer thought that the risk of introducing default methods in replacement of abstract skeletal implementation classes outweighed the benefits because their framework was quite popular and thus extended by many clients. As such, the developers were extremely cautious of introducing widespread syntax (but not semantic) changes throughout their project, especially considering that they were not bug fixes. Indeed, the changes proposed to eclipse/jetty.project comprised one of our largest with nearly 30 modified files. To ease their anxiety, we offered to revert some of the more invasive changes, which were issued in separate commits, such as the declaring class removal. This effort proved to be futile, however.

#### 4.2.3 Choosing a Suitable Default Implementation

**Narrow Implementations**   We found several instances where the migrated (source) methods provided an implementation that was too narrow to be applicable as a default implementation for the particular destination interface. This was in spite of the conservative nature of the refactoring, meaning that the narrow source method implementations were the only ones available that could be migrated to a default method without (i) breaking the rules of such methods (e.g., no field dependence) and (ii) preserving original program semantics. For example, project junit-team/junit5 has a central interface org.junit.platform.engine.TestEngine with a method getId(). Only one implementation, namely, org.junit.jupiter.engine.JupiterTestEngine.getId(), of four was a viable source method as it utilized a constant as a return value while the others relied on instance fields, which cannot be declared in interfaces. However, the value returned in this source method was that of *"junit-jupiter"*, which is too narrow for an arbitrary TestEngine to inherit as a default implementation. To put another way, *"junit-jupiter"* is the name of a *specific* TestEngine, one that would not apply to another implementation and thus would not be considered generally applicable to all potential implementers.

**Cardinality Between Skeletal Implementations and Interfaces**   Enhanced interfaces, for a particular method, allow one and only one skeletal implementation to serve as the default implementation for a method. But, the skeletal implementation pattern allows for *multiple* skeletal implementation classes, each possibly providing their *own* implementation of each method. Implementers can then choose between these by extending the appropriate skeletal implementation class. Furthermore, since default methods become the de facto skeletal implementation for an interface method as they are declared *directly* in the interface, great care should be taken in its selection. A suitable skeletal implementation must be selected to serve as the default for a certain method. The implementation should be beneficial and applicable to *all* potential interface implementers.

We found in at least one case, namely, with the pull request issued to ReactiveX/RxJava, that, despite a unique, viable skeletal implementation existing for each refactored target method, several of the migrated methods would not have been the default implementation of their choosing. This proved to be one of the reasons this pull request was rejected.

**Migrating Skeletal Implementations from Test to Core Packages**   In some cases, as in the migration that occurred with elastic/elasticsearch, interfaces in core packages, i.e., those central to the system and normally having many inward dependencies, had unique skeletal implementation classes emanating from test packages. More than likely, this pattern is used to facilitate easier development of mock objects for testing purposes. However, as tests normally have to deal with artificial data [34, Ch. 8], skeletal implementations used in testing situations are most likely not suitable for

**Proactive Empirical Assessment of New Lang. Feature Adoption via Autom. Refactoring**

default methods for interfaces in core packages. For example, in nhl/bootique, one particular source method of a successful migration was declared in a test package, while the destination interface was not test-related. The implementation was very test-specific and, although it stemmed from a unique skeletal implementation of the interface, it would not have been a suitable default implementation for non-test related implementers. Moreover, as was the case with elastic/elasticsearch, it is possible that default implementations reference elements, e.g., classes, contained in the test package. In these cases, we would have core packages depending on test-based packages, which is architecturally undesirable (e.g., one would *not* want to bundle a test package in a deliverable product but *would* want to bundle a core package).

### 4.2.4 Default Methods and Dependencies
**Primitive Operations** Project aalmiray/jsilhouette accepted our pull request with no modifications or critiques. Although the project is relatively small and not widely used (at least in terms of its activity on GitHub), the change set size, consisting of 147 added lines, 294 removed lines, and 4 changed files, is non-trivial with the total number of changed lines accounting for nearly 20 % of its code base. The average added lines, removed lines, and changed files across *all* projects was 134, 266, and 21, respectively.

Upon further investigation, we noticed that no modification to import statements were included in the change proposal, signifying a self-contained migration. Furthermore, the migrated default method bodies only contain method calls to methods of the destination interface. We speculate that such a use case of default methods is ideal, and further expound in Section 5. In fact, Bloch [3, Item 18] states that candidate methods, so-called *primitive* operations, for skeletal implementations are those implementations that are only in terms of the interface. We conjecture that this notion also extends to default methods. `List.atCapacity()` (lines 5–6, Listing 2b) is an example of such a method, i.e., a default method written only in terms of other (default or non-default) interface methods.

In contrast, the pull request issued to project spring-projects/spring-framework contained a non-trivial amount of import statement changes, which was necessary as the migrated source methods had dependencies outside of the destination interface package. This pull request was rejected, with Clozel [6] citing the following as one of the reasons:

> It is really important for us to keep a clean [code base,] and we have a strict "[zero] package tangle" policy. Moving code (and imports) around is rather risky. I haven't run sonar[qube, a code quality reporting tool [35]] on your [pull request] but I think this might be ...interesting ....

Zero package tangle corresponds to the situation where classes do not form strongly connected components in terms of their package dependencies [13, 15]. Thus, at least this particular project was adverse to the introduction of default methods that had dependencies on external packages, e.g., due to non-interface method calls.

**Constant Fields**   In addition to method declarations and now, with the emergence of Java 8, default method bodies, interfaces are also allowed to declare constant fields. Since interfaces are not allowed to have (variable) state, constants are the only form of fields that an interface can declare. In turn, default methods cannot work with interface state but they *can* work with constants declared in the interface.

Project nhl/bootique, also consisting of an accepted pull request, had two successful migrations where two source methods were not only in terms of solely destination interface methods (cf. Section 4.2.4), but also accessed solely constant fields. In this particular case, the source methods were migrated to the destination interface Environment from the declaring class DefaultEnvironment. The (now default) implementations used two constants whose values consisted of *"bq"* and *"BQ_"*, respectively, to recall environmental variables that began with the given prefix. DefaultEnvironment served the purpose of providing prefixes associated with the entire project. In part, because these constant fields were only accessed by the source methods, the constant declarations could also be safely migrated to the destination interface. Similar to the scenario depicted in Section 4.2.4, this resulted in a self-contained migration and, furthermore, made full use of the (language) capabilities available to interfaces.

**Method Parameters**   We observed that accepted pull requests did not only contain calls to "local" interface methods or references to "local" constant fields but also included calls and references to entities related to method parameters. One such case was observed in the pull request issued to aol/cyclops-react.

**Crossing Build Script Module Boundaries**   Several projects, including eclipse/eclipse-collections, advantageous/qbit, and junit-team/junit5, had build script produced modules separating their APIs (interfaces) from provided implementations, which, perhaps surprisingly, included skeletal implementation (declaring) classes. Thus, migrating source methods from implementation modules to target methods in API modules may introduce dependencies between the modules. This can cause circular dependencies, which are disallowed by many popular build script and/or module systems, e.g., OSGi [1]. For migrations that did not induce circular module dependencies, eclipse/eclipse-collections only accepted those with default methods that were either in terms of other interface methods or did not require external imports.

### 4.2.5 Implicit Reliance on Skeletal Implementation Classes

In spring-projects/spring-framework, there was at least one instance where a (now empty) declaring class was removed because all of its contents were migrated. Additionally, there were apparently no dependencies on the class, and it was deemed safe to remove. However, the class removal caused a test failure for a test dealing with the Aspect-Oriented Programming (AOP) [26] features of the framework. This failure was due to an *indirect* dependency caused by pointcut, i.e., a specification over the program's execution of where code (advice) is to be applied in the underlying program, contained within an aspect. Specifically, the test asserted behavior that would only be present if advice applied to executions of the source method. However, AspectJ [25], a Java AOP implementation, does not consider an interface method as executable if there are *no* classes explicitly implementing the interface. As such, although the source method in question was migrated, for the advice to apply, the empty declaring class had to remain.

#### 4.2.6 Skeletal Implementations and Deprecated Methods

In jOOQ/jOOQ, both several source methods and associated target methods were marked as @Deprecated. The target method javadoc mentioned that different methods of the same (destination) interface should be used instead. The source methods simply delegated (forwarded) calls to the new, replacement methods. With these methods now migrated to the interface, it was clear from the default implementation what the replacement method was, rather than solely relying on documentation for this information. Unfortunately, this pull request was rejected due to the restriction of supporting Java 6 clients and that the developers did not see an overwhelming benefit to the change versus the risk (the change set was rather large with 22 files changed). However, only some of these changes involved deprecated methods, and, as elaborated upon in Section 5.5, deprecated methods are an intriguing use case for default methods, especially considering that documentation may be either absent, incomplete, or outdated.

## 5 Discussion

In this section, we summarize and relate the results in Section 4 with the research questions posed in Section 3.1.

### 5.1 Default Method Adoption R1

Note that, in answering R1, our study focuses solely on how default methods have been adopted in existing code in an effort to replace the classic Skeletal Implementation Pattern. There may be other situations in existing code where default methods can be used but not as a replacement, e.g., introducing a new interface method without breaking existing clients.

#### 5.1.1 Interface Locality

The answer of R1 includes situations where the migrated default methods were self-contained, i.e., that the default implementation was mostly if not fully in terms of *both* methods and constant fields declared either within the same interface or one up its interface hierarchy. In other words, these default methods were composed of either calls to more primitive operations or constant fields in the interface. This typically lent itself to default implementations that were simple and easy to reason about because referenced elements are more or less local to the interface, as well as introduced *no* new external dependencies, thus preserving any architectural and packaging constraints. Circular dependencies between build script modules are also averted.

### 5.1.2 Parameter Locality

Default method implementations that referenced entities related to and called methods on parameters also did not introduce any new dependencies or import statements. Dependencies between the interface and the method parameter types must have existed prior to the default method migration. As such, although parameter types may not have been local to the interface as described above, they were in some respect "local" to the default implementations. Additionally, this is a use case not thoroughly discussed in [3, Item 18] as guidelines for writing skeletal implementations.

### 5.1.3 Optional Interface Methods

eclipse/eclipse-collections accepted several migrations where the default methods simply threw `UnsupportedOperationException`. Default methods seemed useful for these optional methods as implementers can clearly see via the implementation that they are not required to implement them, as well as what should happen if they are not supported. Typically, implementers rely on documentation to make these determinations.

### 5.1.4 Instance Method Interfaces to Static Methods

Also in eclipse/eclipse-collections, we noticed several accepted default methods that were used to provide instance method interfaces to existing static methods. In other words, the default methods allowed clients to call methods on the implementing instance as instance methods that would normally be called as static methods. For example, the default method `ListIterable.binarySearch()` simply forwarded the call to the static method `Collections.binarySearch()`, allowing clients to effectively call the static method as an instance method, which may be more consistent with surrounding code. This typically did not involve introducing new dependencies and required only one call.

## 5.2 Default Method Rejection R2

### 5.2.1 JDK Versions

In answering R2, we encountered several underlying situational themes as to why developers did not wish to incorporate default methods into their projects. The most prevalent, perhaps not surprisingly, was due to backward compatibility restrictions on supporting older Java Development Toolkit (JDK) versions. Java 8 is relatively new, and the popular Android SDK still, at the time of this study, depended on JDK 7. As such, projects that were frameworks were likely to be hesitant in breaking existing Android clients as a result of moving to Java 8. We expect demand for Java 8 features to increase in the near future, especially since they are now (at least partially) supported by Android [21].

The focus of our study is on new language feature adoption. Being restrained by backwards compatibility with earlier platforms is a reason not to adopt new language features. In the case of default methods and Java 8 in particular, the legacy restrictions posed by the widely popular Android SDK is of interest. In deciding to adopt a new language construct, developers must not only consider the language construct itself but also substantial reliance on platform backwards compatibility.

### 5.2.2 Architectural Constraints

Also in addressing R2, developers were not always keen on introducing new external dependencies into interfaces as some default methods required. Doing so sometimes violated both implicit and explicit architectural constraints. For example, as was the case with spring-projects/spring-framework and others, introducing dependencies into modules may deteriorate the architecture and introduce cyclic dependencies. Perhaps more significant is the situation where developers *overriding* the default implementation *still* must provide the code (modules) to satisfy the, e.g., import statements. In this situation, because of default methods, developers may be forced to package modules in their distributions despite not using them in the code.

Another problematic situation is when projects separated their APIs (interfaces) and an implementation of that API into separate modules. In such cases, modules provided by the (framework) developers are *particular* (perhaps reference) implementations of the provided API. If the interfaces are accompanied by code in the form of default methods, (framework) clients providing their *own* API implementations must also package the framework provided (reference) implementation in their distributions. In this situation, despite possibly only one API implementation being utilized, two are delivered to customers, perhaps resulting in bloated deliverables, an added maintenance burden, and decreased program compression. As such, coupling complex default implementations with interfaces may not be practical in these circumstances.

### 5.2.3 Client Enforcement

Several projects, especially those that are frameworks, were particularly anxious regarding "inlining" skeletal implementations directly into interfaces. One of the benefits of default methods is not having to discover and maintain a separate class containing a skeletal interface implementation [24], however, several projects seemed to use this to their advantage. That is, project developers looked to be more comfortable forcing clients to provide an interface method implementation, *despite* providing a skeletal implementation in a separate class. We speculate that having a default implementation front and center by defining it in the interface directly may bring to light any problems it may have and/or that it might not be general enough for all implementers. There was also concern for new (default) interface methods not being overridden by clients that should.

### 5.2.4 The "Correct" Default Implementation

In several cases, we found some skeletal implementations to be too narrow to be the "de facto" implementation of an interface method. The skeletal implementation pattern allows for multiple skeletal implementations per interface method, whereas default methods only allow for one such implementation. Moreover, skeletal implementations were popular in automated testing code and were sometimes the *only* skeletal implementations for a given interface method. However, these implementations are typically not appropriate to be used outside of testing.

### 5.3 Default Method Trade-offs R3

As outlined in Section 4.2.2, there were several cases where developers were anxious regarding the loss of interface implementer control. Compared to the Skeletal Implementation Pattern, default methods are available to *all* interface implementers, regardless whether or not they extend a separate skeletal implementation class. Placing skeletal implementations directly inside interfaces as default methods explicitly presents implementers with a skeletal implementation that they may or may not choose to override with their own. If these implementations are not applicable to implementers, default methods may actually have a negative effect.

One project initially brought up a recently reported performance degradation with default methods as a reason to reject the pull request. However, this was not a deciding factor in the decision as it was seen as a temporary problem with the JVM that affected only a small number of cases.

### 5.4 External Factors R4

Several external factors seemed to influence developer adoption decisions in some cases. Project eclipse/eclipse-collections was significantly the largest project with the largest proposed changeset to accept our pull request. Prior to this study, it had already made relatively extensive use of Java 8 features. Thus, it seemed more likely to adopt default methods, perhaps due to the developers' prior familiarity with Java 8.

As discussed in Section 4.1, smaller changesets were more likely to be accepted than larger ones, although developers had the ability to dissect the refactoring. Nevertheless, the more migrations were from diverse files across module boundaries (evident from column *δ files* in Table 1), the less likely a successful adoption would occur. Lastly, migrations stemming only from abstract skeletal implementation classes were more likely to be adopted, indicating that these source methods were general enough to be applicable to a wider variety of interface implementers.

### 5.5 Best Practices in Using Default Methods R5

In this section, we set forth several best practices and patterns of using default methods based on the experience obtained during this study.

- Default methods should be simple. This reduces the likelihood of introducing complex dependencies into interface modules, creating self-contained default implementations, and having the default implementation serve as an enhancement to the interface documentation (e.g., what optional methods do when called if they are not implemented).
- Take care in using default methods for *new* methods that interface implementers should override. As discussed in Section 4.2.2, default methods may inadvertently mask interface evolution if the developers' intention is to break *existing* imple-

- menters by adding a new interface method so as to alert them to override (implement) new methods.
- Write default methods in terms of (other) methods and constants of the same or closely related interfaces and/or their parameters. Not only can this simplify default method implementations but, as discussed in Sections 4.2.4 and 5.1.2, it can make implementations more self-contained and reduce external dependencies.
- Consider the implications of using default methods during architectural design. As detailed in Section 5.2.2, (particularly framework) developers may need to rethink separating interface declarations and interface implementations into separate modules in light of default methods because they may contain references to implementation modules, which are typically not available to interface modules.
- Forward calls to deprecated interface methods to their replacement API, if applicable, using default methods. As discussed in Section 4.2.6, doing so can enhance method documentation, as well as eliminate any confusion involving deprecation between the interface method and its corresponding skeletal implementation in a separate class.
- Choose default methods that are general enough for *all* potential implementers. As stated in Section 5.3, in contrast to the skeletal implementation pattern, all interface implementers can inherit default implementations *regardless* if they extend any (skeletal implementation) class. In this case, it is essential to choose a general default implementation. Sometimes the only viable option due to language restrictions (cf. Section 4.2.3) is too specific. In these cases, it may be best not to use default methods.

## 6 Threats to Validity

In this section, we consider possible threats that may undermine our study and how they have been mitigated.

**Subject Selection Bias** To minimize bias, none of the authors were involved in the development of any of the subjects prior to conducting the study. Moreover, a wide variety of subjects were selected, varying in size and domain.

**Automated Refactoring Bias** The automated refactoring tool used could be overly conservative in certain situations, and, thus, may not match a human-initiated migration. Furthermore, a human could "massage" the code so that it passes refactoring preconditions, which would be considered too invasive for the automated tool to mimic. Therefore, the refactoring produced by the tool may vary from what an expert developer may produce, which could influence developers' decisions. However, since the refactoring tool is highly conservative, the changes made are minimal; developers were free to alter the code further. In other words, the conservative nature of the refactoring used in the study can be seen as a bridge to initiate a more large-scale or customized refactoring. As such, the refactoring issued by an automated tool could be used as a guide of possible locations where default methods may be applicable. Moreover, using the results of the tool as a guide, developers had the option of choosing a different implementation by altering the source code directly as necessary. Lastly, developers were allowed to critique the automatic refactoring, as well as request changes, which, while mostly minor, were accommodated.

**Simulating Human Development**  When developers introduce a new language feature manually, it may be done in step-wise fashion rather than in bulk. Henceforth, perhaps large changes deterred developers from accepting our pull requests due to risk, especially when the project included a popular external API. Other projects, e.g., google/binnavi, cited concerns on the change set size and that they preferred to apply the refactoring in a more incremental fashion as opposed to merging (accepting) a large pull request. However, many of the developers, despite being introduced to large change sets, carefully reviewed pull requests in a step-wise fashion, inspecting and often commenting on smaller changes. This behavior, in a sense, is approximate to a step-wise process. To further mitigate this threat, our study was comprised of a wide range of change set sizes, both in terms of number of altered lines of code and files. We also typically demarcated commits so that they separately included the default method migration, i.e., where the migration occurs, removal of declaring classes, and javadoc modifications. This afforded developers a path of step-wise changes to follow so that they may consider the change impact more in depth.

**Tool vs. Construct Assessment**   It could be argued that acceptance or rejection of pull requests may not equate to acceptance or rejection of the default method construct. However, since this work is focused on the introduction of default methods into existing code, i.e., that the introduction of the construct results in semantically-equivalent code, and that replacing the skeletal implementation pattern is the only sensible motivation of using default methods without extending interfaces [17], acceptance and/or rejection of pull requests due to the refactoring result contents must equate to the acceptance and/or rejection of the construct itself. This is also due to the conservative, unambiguous, and fully-automated refactoring tool used during the study. The developer discussion presented in the prior section holds as evidence of these facts.

**Open Source Software**  Our study deals with the adoption of default methods in primarily open source software. However, results of studying open source software may not generalize to all software development scenarios. In particular, there may be characteristics and peculiarities that are specific to open source development that may not be found in, e.g., a commercial product. It could be argued that closed source or commercial software developers would be more reluctant to take risks compared to open source developers since their customers are paying for their product. However, a counter argument could be made that, since open source software is publicly available, it has a wider audience and thus any potential breakage would have a comparable, if not greater, impact. In fact, many private companies use, invest in, and contribute to open source software.

## 7 Related Work

### 7.1 Investigating Language Features and Library Usage

Others have also inquired into how developers use new language features. For instance, Parnin, Bird, and Murphy-Hill [31] study the usage of Java generics, particularly, whether they meet their intended purpose of relieving explicit type casting, measuring their effectiveness and adoption by automatically mining open source project histories. Dyer, Rajan, Nguyen, and Nguyen [14] perform a similar but more general study. Their work studies different language features, both in terms of usage and intended benefits. Also, their analysis is postmortem while ours is proactive, enabling us to study language features much sooner after their initial release. They also do not provide usage best practices.

Uesbeck, Stefik, Hanenberg, Pedersen, and Daleiden [40] assess the human factor impact of C++ lambda expressions on developers by comparing their usage to that of iterators. They do so via a study of how long students spent writing programs with lambda expressions. This approach is concerned with ascertaining the learning curve associated with the new language feature during programming tasks. Hoppe and Hanenberg [19] perform a similar study but for Java generics. Our approach, on the other hand, enables the study of reactions of experienced developers in incorporating a new language feature into their existing projects that may be used by many other experienced developers and whether the feature improves their projects in some way.

Wu, Chen, Zhou, and Xu [41] examine how C++ concurrency constructs are used in open source software by examining the code with no developer input. Gorschek, Tempero, and Angelis [18] and Tempero, Counsell, and Noble [38] inspect the usage of other Object-Oriented language features, namely, information hiding and encapsulation and inheritance, using surveys and metrics, respectively. Souza and Figueiredo [36] scrutinize the use of optional types in dynamic languages. Lin, Okur, and Dig [27] investigate how developers retrofit asynchrony into Android apps, while Okur, Hartveld, Dig, and Deursen [29] scrutinize usage of programming constructs for asynchronous programming via a large scale study of C# programs. Okur and Dig [28] analyze the usage of parallel libraries in a large scale study. Each of these involves a level of postmortem analysis.

### 7.2 Mining Pull Requests

Rahman and Roy [33] also analyze GitHub pull requests but for the purpose of gaining insights into why and how they are either accepted or rejected. Our work, on the other hand, uses pull requests for developer feedback on a new language construct. They also study developer conversations inside of pull request bodies (discussion text) and project information, and go beyond our methodology to further examine developer project maturity and specific developer information such as experience. It is conceivable that our work could also benefit from project maturity and domain dimensions to our analysis. However, since our technique relies on the output of an automated refactoring tool, developer experience would most likely be less useful.

### 7.3 Refactoring Automation

Khatchadourian and Masuhara [23] present a default method refactoring approach and corresponding evaluation, the implementation of which is utilized in this work. While Khatchadourian and Masuhara [23] bring forth the results of a brief preliminary pull request study, it is solely focused on assessing the accuracy of the refactoring. The current work presented in this paper, on the other hand, is *not* an assessment of the refactoring tool but *rather* a presentation and study of a technique where an automated refactoring is used to *ascertain the usefulness of default methods*. Here, we include thorough details of developer reactions, which include situations where pull requests were merged and where they were rejected, as well as outline the reasons for each. We also set forth best practices in using default methods.

## 8 Conclusion & Future Work

We describe a novel proactive approach, using automated refactoring, to empirically assess new programming language features early. The new construct is introduced to developers as refactoring results in which they decide whether to incorporate, regardless of any previous experience, and at the same time providing insight into their decisions. This facilitates reasons why new features are not adopted as such cases may not be explicitly documented, thus possibly alluding traditional postmortem approaches. Moreover, the developers, typically project committers, being questioned are experienced and thus more than likely to provide feedback that is superior in quality over traditional student studies.

Our approach was applied to 19 open source projects to assess Java 8 default methods. Scenarios where and reasons why default method migrations from the skeletal implementation pattern were either accepted or rejected by developers were put forth, and best practices were extracted. This insight can not only benefit developers but also language designers, especially those considering similar constructs for other languages.

In the future, we plan to apply our proactive approach in studying other new language features such as those available as part of Project Jigsaw [12], which will be part of Java 9 [11]. Additionally, now that Android supports Java 8 features [21], we plan to conduct a specialized study of new language features in the context of mobile applications.

Rather than investigating adoption of *new* language features, in the future, we plan to apply our approach to investigating those that are currently being *proposed*. The goal would be to justify the introduction of a possible future language feature being considered. Doing so would provide further insight into the usefulness of the proposed feature prior to its wider deployment. With the proactive method presented here, language designers may create an automated refactoring along with the language feature implementation and apply both to existing projects. The results may then be submitted as pull requests to project developers for proactive feedback on the proposed feature.

Another avenue of interesting future work is to examine the relationship between refactored code and unit test case coverage. Test coverage may serve as an external factor for developers in deciding if the proposed change is too risky and conversely may be more likely to accept a pull request if the code changes are well-covered by existing test cases.

Lastly, our study results may be used to enhance the Migrate Skeletal Implementation to Interface refactoring tool. Specifically, the knowledge gained may be formulated to prioritize and/or rank refactoring suggestions. Suggestions are normally presented by file. An alternate approach inspired by the results of this study would be to present the suggestions so that the ones more likely to be desired by the developer would be presented first. To achieve this, a cross-examination can be made between the change set proposal and general heuristics derived from accepted pull requests, e.g., refactorings involving fewer introduced dependencies could be ranked higher than ones with more.

**Acknowledgements** This material is based upon work supported by PSC-CUNY under award #69165-00 47 and the Tokyo Institute of Technology Research Abroad program.

## About the authors

**Raffi Khatchadourian** is an Assistant Professor in the Department of Computer Science at Hunter College and Doctoral Faculty of the Graduate Center PhD Program in Computer Science of the City University of New York. His research interests are primarily in the software engineering and programming languages fields, particularly in regards to automated software evolution. Contact him at raffi.khatchadourian@hunter.cuny.edu or rkhatchadourian@gc.cuny.edu.

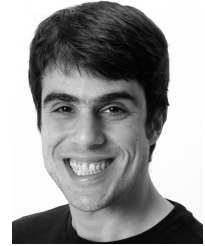

**Hidehiko Masuhara** is a Professor of Mathematical and Computing Science at Tokyo Institute of Technology. His research interest is programming languages, especially on aspect- and context-oriented programming, partial evaluation, computational reflection, meta-level architectures, parallel/concurrent computing, and programming environments. Contact him at masuhara@acm.org.

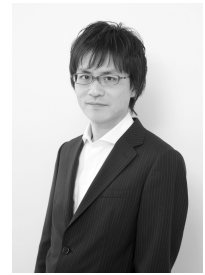